\documentclass[10pt]{article}
\usepackage{amsmath}
\usepackage{amssymb}
\usepackage{hyperref}
\usepackage{graphicx}
\usepackage{epstopdf}
\usepackage{cite}
\usepackage{color}
\usepackage{filecontents}
\usepackage[labelfont=bf,labelsep=period,justification=raggedright]{caption}
\makeatletter
\renewcommand{\@biblabel}[1]{\quad#1.}
\makeatother

\date{}
\begin{document}
% Title must be 150 characters or less
\begin{flushleft}

{\Large \textbf{Attention Dynamics in Collaborative Knowledge Creation} }
% Insert Author names, affiliations and corresponding author email.
\\
Lingfei Wu$^{1,\ast}$, Marco A. Janssen$^{1,2}$
\\
\bf{1} Center for Behavior, Institutions and the Environment, Arizona State University, Tempe, AZ 85281, U.S.
\\
\bf{2} School of Sustainability, Arizona State University, Tempe, AZ 85281, U.S.
\\
$\ast$ E-mail: Lingfei.Wu@asu.edu
\end{flushleft}

% Please keep the abstract between 250 and 300 words
\section*{Abstract}

To uncover the mechanisms underlying the collaborative production of knowledge, we investigate a very large online Question and Answer system that includes the question asking and answering activities of millions of users over five years. We created knowledge networks in which nodes are questions and edges are the successive answering activities of users. We find that these networks have two common properties: 1) the mitigation of degree inequality among nodes; and 2) the assortative mixing of nodes. This means that, while the system tends to reduce attention investment on old questions in order to supply sufficient attention to new questions, it is not easy for novel knowledge be integrated into the existing body of knowledge. We propose a mixing model to combine preferential attachment and reversed preferential attachment processes to model the evolution of knowledge networks and successfully reproduce the observed patterns. Our mixing model is not only theoretically interesting but also provide insights into the management of online communities. 

%\vspace{20 mm}

\section*{Introduction}
Knowledge is created at the cost of human effort, in particular, the collective attention of contributors. There are two types of tasks involved in knowledge creation, one is to further process existing knowledge and the other is to explore emerging topics. As both of them consume collective attention, which is a limited resource \cite{simon1973applying}, it is important to balance between them so that our society can be benefit from the attention investment continuously. However, this goal is becoming more challenging due to the conflict between information overload and attention scarcity in the contemporary world \cite{wu2007novelty,weng2012competition,lehmann2012dynamical,wu2013metabolism,wu2013decentralized}. 

To investigate how real-world knowledge production systems balance between the fine graining of old knowledge and the exploration of new knowledge, we collect and analyze the historical data of StackExchange. Launched in Sep., 2009, StackExchange is currently one of the largest question and answer systems in the world, containing many communities on topics in varied fields. For each of the 110 communities in our dataset, we extract a knowledge network in which nodes are questions and edges are the answering activities of users that connect sequentially answered questions. As questions reflect the demand of knowledge and answers represent the supply of attention, the linking structure of this network thus reveals the strategy of the studied system in allocating attention (edges) on knowledge (nodes). 

We find that the knowledge networks have two properties: 1) the mitigation of degree inequality among nodes. More specifically, the distribution of degree is not as skewed as predicted by the classical preferential attachment model \cite{barabasi1999emergence}. As a consequence, the highest degree, i.e., the maximum number of answers collected by a single question, increases slowly with network size; 2) the assortative mixing of nodes. High-degree nodes tend to connect with each other, leading to an increase of neighbor connectivity (the average degree of the neighbors) with degree \cite{pastor2001dynamical,newman2002assortative} and the separation of high-degree nodes from low-degree nodes. These findings, put together, provides insight into the collaborative knowledge production process: while the system is trying to reduce attention investment on old questions in order to supply sufficient attention to new questions, it is still not easy for the novel knowledge created during this process be integrated into the existing cluster of old, well-discussed body of knowledge. 

Based on this analysis, we assume that there are two mechanisms in the system. The first one is to supply attention preferentially to popular questions that already have many answers, so that existing knowledge will be further processed. The other one is to supply attention in prior to questions that have not attracted many answers, so that emerging topics can be addressed promptly and the scope of knowledge is continuously extended. We model the first mechanism by preferential attachment (PA) \cite{barabasi1999emergence}, in which the attractiveness of an existing node to new nodes increases with its degree, and the second mechanism by the ``reversed" preferential attachment (RPA), in which attractiveness decreases with degree. A mixing ratio $p$ ($0\leq p \leq 1$) is used to control the mixture between these two linking dynamics during the evolution of knowledge networks. Our model successfully reproduces the two observed properties of knowledge networks.

\section*{Materials and Methods}

StackExchange is a network of question and answer communities covering diverse topics in many different fields. We downloaded its database dump on Jan., 2014 from a publicly accessible archive (https://archive.org/ details/stackexchange), which contains the log files of 110 communities since their launch date to the date of data collection. The duration and size vary a lot across these communities. For example, the smallest community on Italian language (italian.stackexchange.com) was created in Nov, 2013 and has 374 users, 194 questions, and 387 answers; while the largest community on programming languages (stackoverflow.com) was created in Jul., 2008 and has 2,728,224 users, 6,474,687 questions and 11,540,788 answers.

For each of the communities we construct a knowledge network including questions as nodes and answers as edges. More specifically, we trace the daily answering activities of a user, and connect the sequentially answered questions one by one following his/her answering order. Aggregating these daily individual answering streams over the period of observation (varying from one to five years, depending on the community under study) gives a directed, weight network that includes the information on the supply (answer edges) and demand (question nodes) of knowledge. For the current study we only focus on the undirected, un-weighted linking structure of the network. 

\section*{The Divided Knowledge }

\begin{figure*}[ht]
\begin{center}
\centerline{\includegraphics[width=0.6\textwidth]{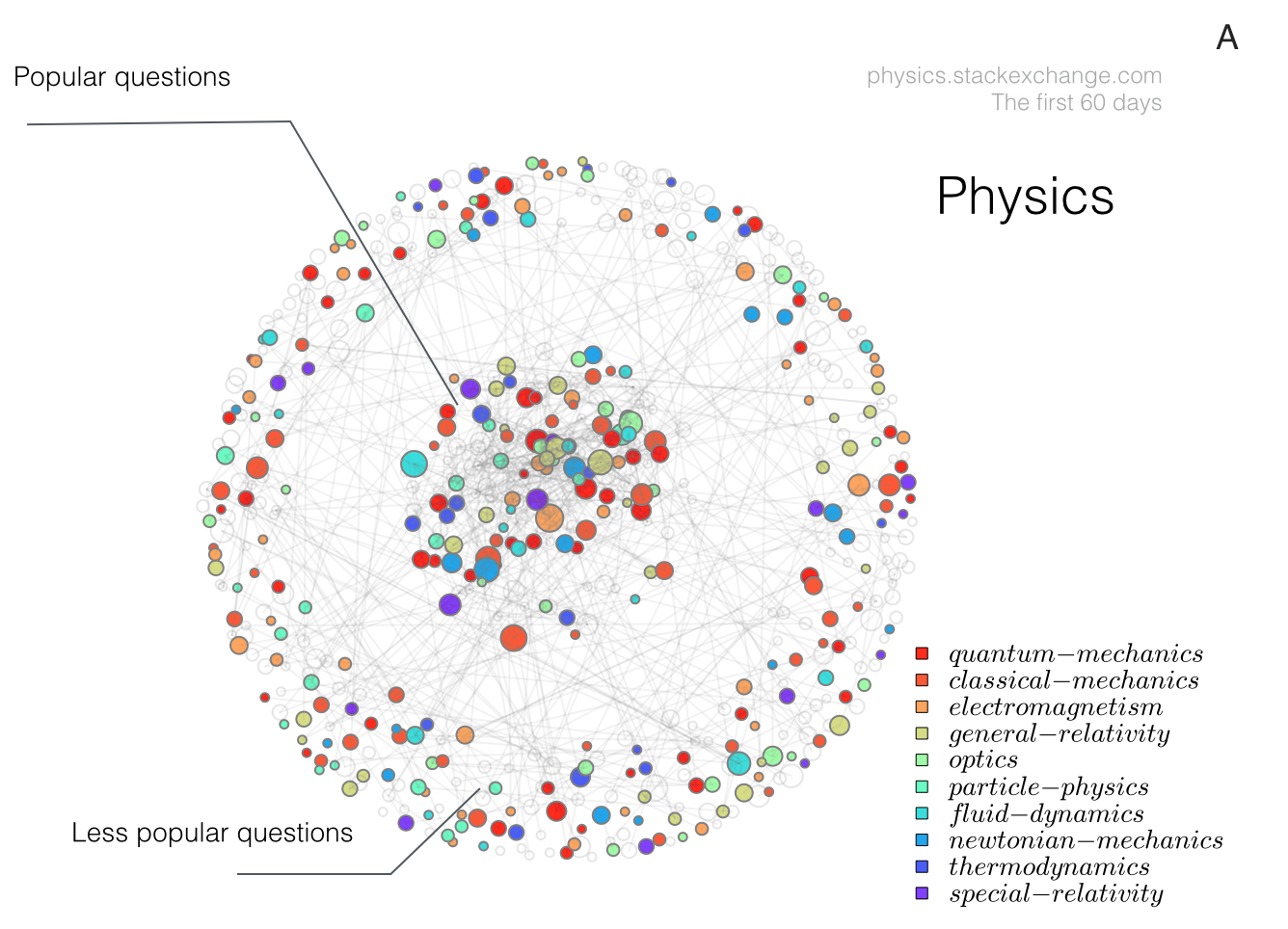}
\includegraphics[width=0.6\textwidth]{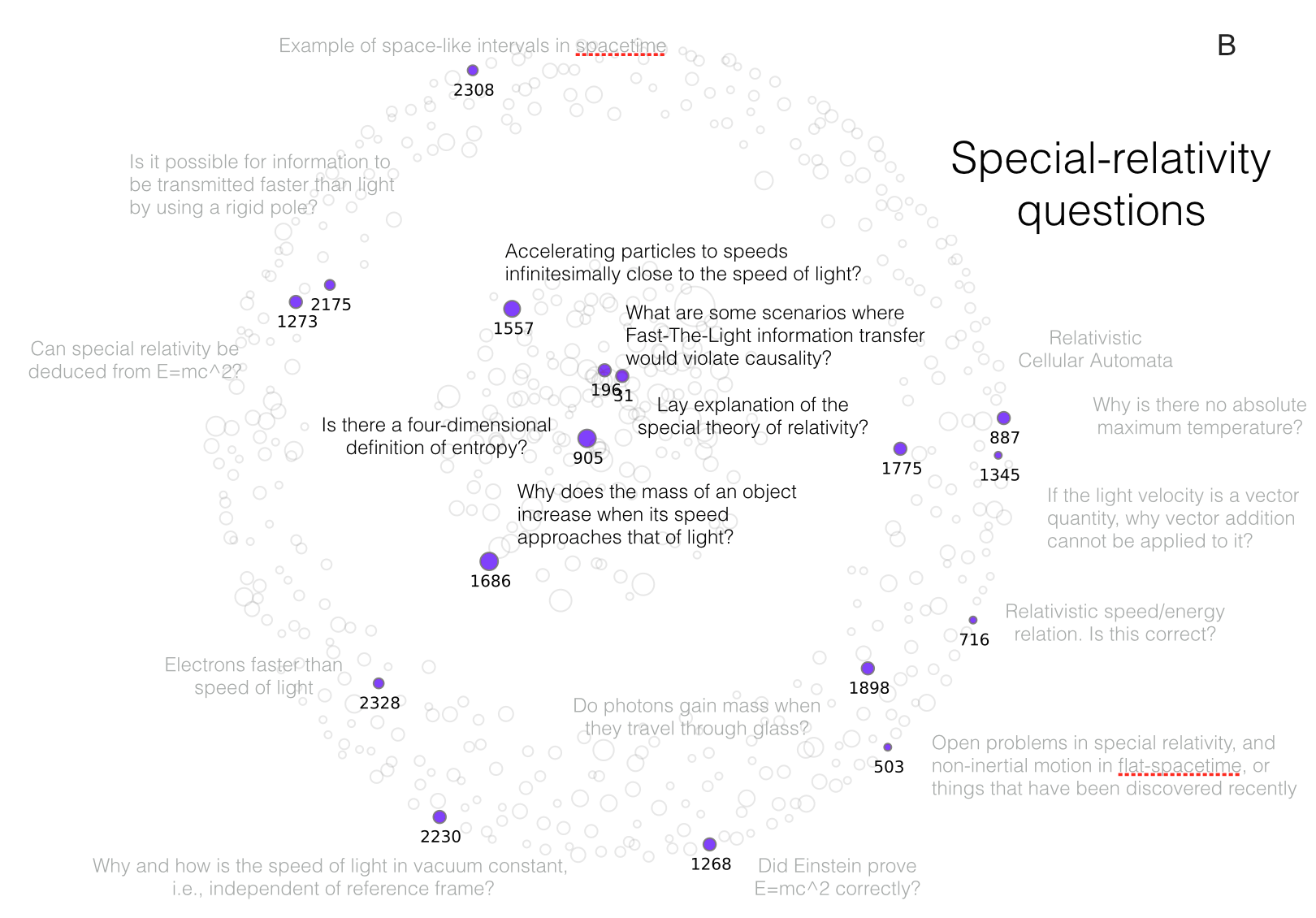}}
\caption{The knowledge network constructed using the 60-day data of physics.exhcange.com, a community for questions in physics in the StackExchange system. In Panel A the nodes are questions and the edges represent the successive answering activities of users. The categories of questions are detected using the first tag of questions and are colored coded. To obtain an visualization that shows the ``distance" between questions (nodes) we calculate the average path length $L_{ij}= L_{ji}$ between all pairs of nodes based on the method introduced in [14], and use the simulated annealing method to embed this network into a two-dimensional Euclidean space such that the total sum of squares of the differences between two-dimensional Euclidean distance and $L_{ij}$ over all pair of nodes is minimized. As shown by Panel A, the core of the network, which is a cluster of high-degree nodes, is separated from the periphery formed by low-degree nodes. Note that due to a mitigation effect of degree inequality discussed in the text, the degree difference between these two parts is observable, but not so apparent. See Figure \ref{figure3}B for the systematic analysis on the assortativity of this network after 5.5 years. In Panel B we display the text of questions relevant to the special-relatively theory. It is observed that the core cluster includes popular questions, whereas the periphery contains specific or trivial questions that fail to generate general interest.
}
\label{figure1}
\end{center}
\end{figure*}

To obtain an intuitive understanding of the linking structure of the knowledge networks we use a novel visualization technique as shown in Figure \ref{figure1}. For a network, we firstly calculate the average length of all possible paths $L_{ij}= L_{ji}$ between each pair of nodes using the method introduced in \cite{wu2015hidden}. Then we use the simulated annealing method to embed the network into a two-dimensional Euclidean space. The target of simulated annealing is to minimize the total sum of squares of the differences between Euclidean distance and $L_{ij}$ over all pairs of nodes. The upper bound of $L_{ij}$ is set to be the diameter of the embedding space to control unreachable nodes. As this visualization technique encodes the average path length between nodes and displays it as Euclidean distance, it works better than force-directed layouts \cite{kobourov2012spring} (which only consider the directed links between nodes) in terms of showing the hidden, global structure of networks.

We apply this technique to the Physics network (the network constructed using the data of \url{physics.exchange.com}) after the first 60 days and obtain Figure \ref{figure1}. It is observed that this network is separated into two parts. The core of the network consists of general, popular questions of many answers (high-degree nodes), such as ``lay explanation of the special theory of relativity?" or ``accelerating particles to speeds infinitesimally close to the speed of light", whereas the periphery is formed by specific, less popular questions (low-degree nodes), such as ``relativistic cellular automata" or ``did Einstein prove $E=mc^2$ correctly?". Note that there could be different reasons why the periphery questions do not get attention. Some of them, like the ``relativistic cellular automata" question, are worthy of exploration but require very specific expertise to answer. But some are simply too naïve or too trivial to generate interest among the community, such as the question ``Did Einstein prove $E=mc^2$ correctly?".    

\section*{Modeling the Allocation of Attention on Questions }

Based on the observation on the separation between old and new knowledge, we assume that there exist two different mechanisms governing the supply of attention. One mechanism is to supply attention preferentially to popular questions that already have many answers. This mechanism contributes to the production of fine-grained knowledge \cite{guan2014fine} as well as the consensus within the community on existing knowledge. The other mechanism is to supply attention in prior to less popular questions that have not been discussed extensively. The existence of this mechanism guarantees that the emerging issues can be addressed promptly, so that the body of knowledge can have a sustainable growth. 

\begin{figure*}[ht]
\begin{center}
\centerline{\includegraphics[width=0.5\textwidth]{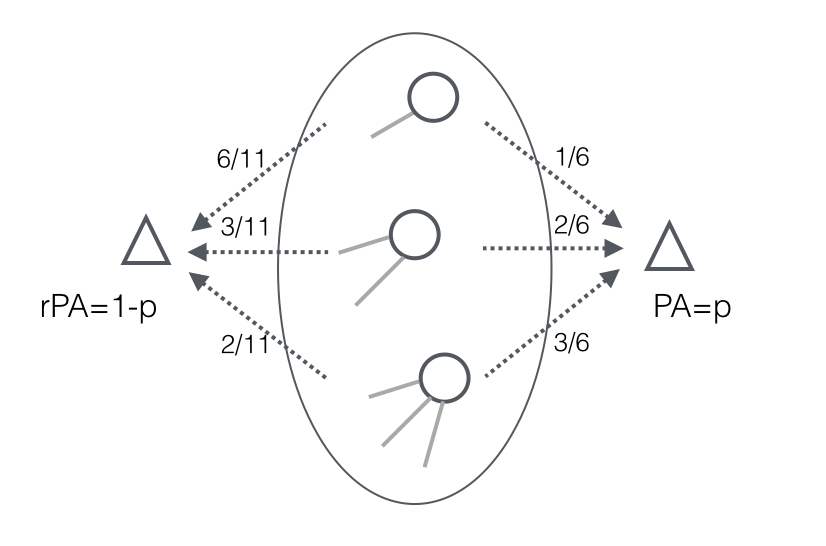}}
\caption{A schematic diagram that shows the mixing between preferential attachment (PA) and reversed preferential attachment (RPA). The nodes are questions and the links are answers that connect sequentially answered questions. The circles represent existing questions in the system and the triangles show the new questions to be added into the system. The probability of a new node be connected to an existing node is marked on the dotted arrow lines. 
}
\label{figure2}
\end{center}
\end{figure*}

The co-existence of these two mechanisms can be modeled by a mixing of two different linking dynamics during the growth of knowledge networks. The mechanism that favors old, popular questions leads to preferential attachment (PA) \cite{barabasi1999emergence}, in which the attractiveness of an existing node to new nodes is directly proportional to its degree; and the mechanism that favors new, less popular questions results in the ``reversed" preferential attachment (RPA), in which the attractiveness is inversely proportional to degree. A schematic diagram of the discussed mixed linking dynamics is shown in Figure \ref{figure2}.

In the following sections, we will perform analysis on our mixing network model to derive its stable degree distribution, the relationship between highest degree and network size, and the assortative or disassortative mixing of nodes. We will begin our analysis by discussing the RPA model alone, which is observed in limited-resource systems \cite{dunne2002food,sevim2006effects} but has not yet been discussed extensively. 

\subsection*{The RPA Model}

The classical preferential attachment (PA) model \cite{barabasi1999emergence} has been shown to successfully describe the centralization of edges on earlier nodes in social \cite{mislove2013empirical}, biological \cite{eisenberg2003preferential}, and technological \cite{capocci2006preferential} networks, but fail to explain why we also observe new nodes beat old nodes and acquire more edges. To overcome this limitation, factors like fitness \cite{bianconi2001competition} and aging \cite{wang2013quantifying} have been added into the model to obtain revised versions of PA.

The systems that can not be explained by the original version of PA model, such as food webs \cite{dunne2002food,sevim2006effects} and citation networks \cite{wang2013quantifying} , usually involve the allocation of a limited resource. The PA model assumes that the attractiveness always increases with degree. This is not true when links have a cost and nodes have a limited capacity thus can not afford a overload of links. For example, in food webs, if there are already too many predator nodes feeding on the same prey node, this prey node may not be so attractive to other predator nodes \cite{sevim2006effects}. Similar to how species compete for energy in ecological systems, information pieces also compete for the collective attention in social systems \cite{weng2012competition}. If a question has already attracted many answers, some answerers many want to avoid the competition and try to contribute to less popular questions. This kind of behavior will equalize edges among nodes and generate many connections between low-degree nodes during the growth of knowledge networks. 

\subsubsection*{The Stable Degree Distribution of the RPA Model}

To model the discussed mechanism that favors less popular questions, we consider an undirected network in which the probability of connecting to an existing node $i$ is proportional to $k{_i}^{-1}$. A similar process was mentioned in ecological studies \cite{sevim2006effects}, but here we present a different version of analytical work to derive stable degree distribution. Assuming at every time step, a new node carrying $m$ links joins the network and connect to existing nodes at rate $R_k$, a function that only depends on the degree $k$ of existing nodes, the average degree of the $s$-th node at time $t$ can be computed from the rate equation
\begin{equation}
\label{eq.1}
\frac{dk_s(t)}{dt}=mR_k=\frac{\frac{m}{K_s(t)}}{\sum\limits_{i=1}^t\frac{1}{k_s(t)}}\approx\frac{\frac{m}{K_s(t)}}{\sum\limits_{i=1}^t\frac{1}{<k_s(t)>}}=\frac{\frac{m}{K_s(t)}}{\frac{t}{2m}}=\frac{2m^2}{tk_s(t)}.
\end{equation}
Rearrange this equation and take the indefinite integral of both sides, we get
\begin{equation}
\label{eq.2}
\int k_s(t)dk_s(t)=2m^2\int\frac{1}{t}dt.
\end{equation}
Considering the boundary condition $k_s(s)=m$ leads to
\begin{equation}
\label{eq.3}
k_s(t)^2=m^2[4ln(\frac{t}{s})+1].
\end{equation}
Eq. \ref{eq.3} predicts the average degree of the $s$-th node at time $t$. As earlier nodes always have higher degrees, we can derive the cumulative probability function
\begin{equation}
\label{eq.4}
P\{k_i\leq k\}=1-\frac{s}{t}\approx1-e^{\frac{1}{4}-\frac{k^2}{4m^2}}\,\,\,\,(k\geq m),
\end{equation}
or the probability function
\begin{equation}
\label{eq.5}
P(k)\approx\frac{k}{m}e^{\frac{1}{4}-\frac{k^2}{4m^2}}\,\,\,\,(k\geq m).
\end{equation}
Our analysis shows that $P(k)$ decays faster than exponential functions with $k$. This is because the RPA process keeps equalizing edges over nodes, thus it is very unlikely to find high degree nodes in the system. 

\subsubsection*{Highest Degree vs. Network Size in the RPA Model}

We can also derive the relationship between highest degree and network size from Eq. \ref{eq.4}. For the highest degree we assume that there is only one node whose degree is greater or larger than this node in the network, which gives
\begin{equation}
\label{eq.6}
N(1-P\{k_i\leq k_{max}\})=1.
\end{equation}
Substituting Eq. \ref{eq.4} into Eq. \ref{eq.6} gives 
\begin{equation}
\label{eq.7}
k_{max}\sim\sqrt {ln(n)}.
\end{equation}

\subsection*{The Model Mixing PA and RPA mechanisms}

In the last section we discussed the RPA process, which is theoretically interesting but is unlikely to be the single rule that governs the evolution of knowledge networks. It is reasonable to assume that there are also users seek for popular questions and prefer to generate fine-grained knowledge. To obtain a more realistic model we combine PA and RPA processes together with a mixing ratio $p$ ($0\leq p \leq 1$).

\subsubsection*{The Stable Degree Distribution of the Mixing Model}

 The increase of the average degree of nodes over time is described by
\begin{equation}
\label{eq.8}
\frac{dk_s(t)}{dt}=m[pB_k+(1-p)R_k]\approx m[p\frac{k_s(t)}{2mt}+(1-p)\frac{2m}{tk_s(t)}].
\end{equation}
in which $B_k$ and $R_k$ represent the increasing rate of new links following the PA and the RPA mechanisms, respectively. Rearrange this equation and take the indefinite integral of both sides, we get 
\begin{equation}
\label{eq.9}
\int\frac{2k_s(t)}{pk_s(t)^2+4m^2(1-p)}dk_s(t)\approx\int\frac{1}{t}dt.
\end{equation}
Considering the boundary condition $k_s(s)=m$ leads to
\begin{equation}
\label{eq.10}
k_s(t)^2\approx\frac{m^2}{p}(4-3p)(\frac{t}{s})^p-\frac{4m^2(1-p)}{p},
\end{equation}
or approximately,
\begin{equation}
\label{eq.11}
k_s(t)\sim(\frac{t}{s})^\frac{p}{2},
\end{equation}
Eq. \ref{eq.11} predicts the average degree of the $s$-th node at time $t$. From Eq. \ref{eq.11} we can derive cumulative probability distribution
\begin{equation}
\label{eq.12}
P\{k_i\leq k\}\sim k^{-\frac{2}{p}}
\end{equation}
or the probability function
\begin{equation}
\label{eq.13}
P(k)\sim k^{-(1+\frac{2}{p})}.
\end{equation}
Eq. \ref{eq.13} predicts that the tail of degree distribution can be fitted by a power-law function of exponent $\alpha=1+2/p \geq 3$. When we increase the value of $p$, the probability of the PA process, the value of $\alpha$ decreases and the degree distribution becomes more skewed. When $p=1$, Eq. \ref{eq.13} degenerates to a power-law function of exponent $\alpha=3$ as predicted by the classical PA model \cite{barabasi1999emergence}.  These predictions are supported by the empirical analysis presented in Figure \ref{figure3}A and network simulation results shown in Figure \ref{figure3}F.

\subsubsection*{Highest Degree vs. Network Size in the Mixing Model}

The relationship between highest degree and network size when $p=1$ can be derived using the similar approach introduced in Section 4.1.2 as
\begin{equation}
\label{eq.14}
k_{max}\sim\sqrt{n}.
\end{equation}
Comparing Eq. \ref{eq.14} with Eq. \ref{eq.17}, we predicts that in data the highest degree should increase with network size faster than $ \sqrt{ ln(n)}$ and slower than $ \sqrt{n}$. This prediction is supported by Figure \ref{figure3}C.

\subsubsection*{The Assortativity of the Mixing Model}

We can also calculate the neighbor connectivity, or the average degree of the neighbors, as a function of degree as follows. We define $R_s(t)$ as the sum of the degrees over the neighbors of node $s$, evaluated at time $t$. That is,
\begin{equation}
\label{eq.15}
R_s(t)=\sum\limits_{j=1}^{k_s(t)}k_j(t).
\end{equation}
The neighbor connectivity $<k_{nn}>$ can be calculated as $R_s(t)/k_s(t)$. During the growth of the network, $R_s(t)$ can only increase by the connection of a new node either directly to node $s$, or to one of its neighbors $j$. In the first case $R_s(t)$ increases by $m$, and in the second case it increases by one unit \cite{barrat2005rate}. Thus, the rate equation of $R_s(t)$ can be expressed as 
\begin{equation}
\label{eq.16}
\frac{dR_s(t)}{dt}=m[m[pB_k+(1-p)R_k]] + \sum\limits_{j=1}^{k_s(t)}m[pB_j+(1-p)R_j],
\end{equation}
which also reads
\begin{equation}
\label{eq.17}
\frac{dR_s(t)}{dt}=\frac{mp}{2t}k_s(t)+\frac{2m^3(1-p)}{t}\frac{1}{k_s(t)}+\frac{p}{2t}\sum\limits_{j=1}^{k_s(t)}k_j(t)+\frac{2m^2(1-p)}{t}\sum\limits_{j=1}^{k_s(t)}\frac{1}{k_j(t)}.
\end{equation}
Substituting Eq. \ref{eq.15} into Eq. \ref{eq.17}, and assuming 
\begin{equation}
\label{eq.18}
\sum\limits_{j=1}^{k_s(t)}\frac{1}{k_j(t)}\approx\sum\limits_{j=1}^{k_s(t)}\frac{1}{<k_j(t)>}=\frac{k_s(t)}{R_s(t)},
\end{equation}
we get 
\begin{equation}
\label{eq.19}
\frac{dR_s(t)}{dt}=\frac{mp}{2t}k_s(t)+\frac{2m^3(1-p)}{t}\frac{1}{k_s(t)}+\frac{p}{2t}R_s(t)+\frac{2m^2(1-p)}{t}\frac{k_s(t)}{R_s(t)}.
\end{equation}
Substituting Eq. \ref{eq.11} into Eq. \ref{eq.19} leads to
\begin{equation}
\label{eq.20}
\frac{dR_s(t)}{dt}=\frac{p}{2t}R_s(t)+\frac{mp}{2}s^{-\frac{p}{2}}t^{\frac{p}{2}-1}+\frac{2m^3(1-p)}{t}\frac{1}{k_s(t)}+\frac{2m^2(1-p)}{t}\frac{k_s(t)}{R_s(t)}.
\end{equation}
It is difficult to derive analytical solution for Eq. \ref{eq.20}. But as our task is to analyze the change of $<k_{nn}>=R_s(t)/k_s(t)$ with $k_s(t)$, we can simply compare the increasing speed between $R_s(t)$ and $k_s(t)$ over time to see whether $<k_{nn}>$ is an increasing or decreasing function of $k_s(t)$. We perform a naïve analysis as follows. Noticing that all items in the right hand side are positive and when $p$ approaches 1, we only need to keep the first two items, which greatly simplifies Eq. \ref{eq.20} as
\begin{equation}
\label{eq.21}
\frac{dR_s(t)}{dt}+(-\frac{p}{2t}R_s(t)) \geq \frac{mp}{2}s^{-\frac{p}{2}}t^{\frac{p}{2}-1}
\end{equation}
Using the integrating factor technique we derive
\begin{equation}
\label{eq.22}
R_s(t) \geq \frac{mp}{2}(\frac{t}{s})^{\frac{p}{2}}[ln(t)+c_1]+c_2 \geq \frac{mp}{2}k_s(t)ln(t),
\end{equation}
in which $c_1$ and $c_2$ are the constants of integration. Therefore, when $p$ approaches 1, $<k_{nn}>=R_s (t) / k_s (t)$ is a constant increasing with system size $t$ . When $p$ is smaller than 1, considering the last two terms in Eq. \ref{eq.20}, we know that $<k_{nn}>$ increases with $k_s (t)$. Meanwhile, the smaller the value of $p$ is, the faster $<k_{nn}>$ increases with $k_s (t)$. In sum, the mixing model is assortative, and this assortativity is negatively correlated with the value of $p$, as evident in Panel B and E in Figure \ref{figure3}.

\section*{The Empirical Data of StackExchange}

We analyze the collected 110 knowledge networks and find that the mixing model explains the observed patterns better than the PA model or the RPA model alone. 

\begin{figure*}[ht]
\begin{center}
\centerline{\includegraphics[width=1\textwidth]{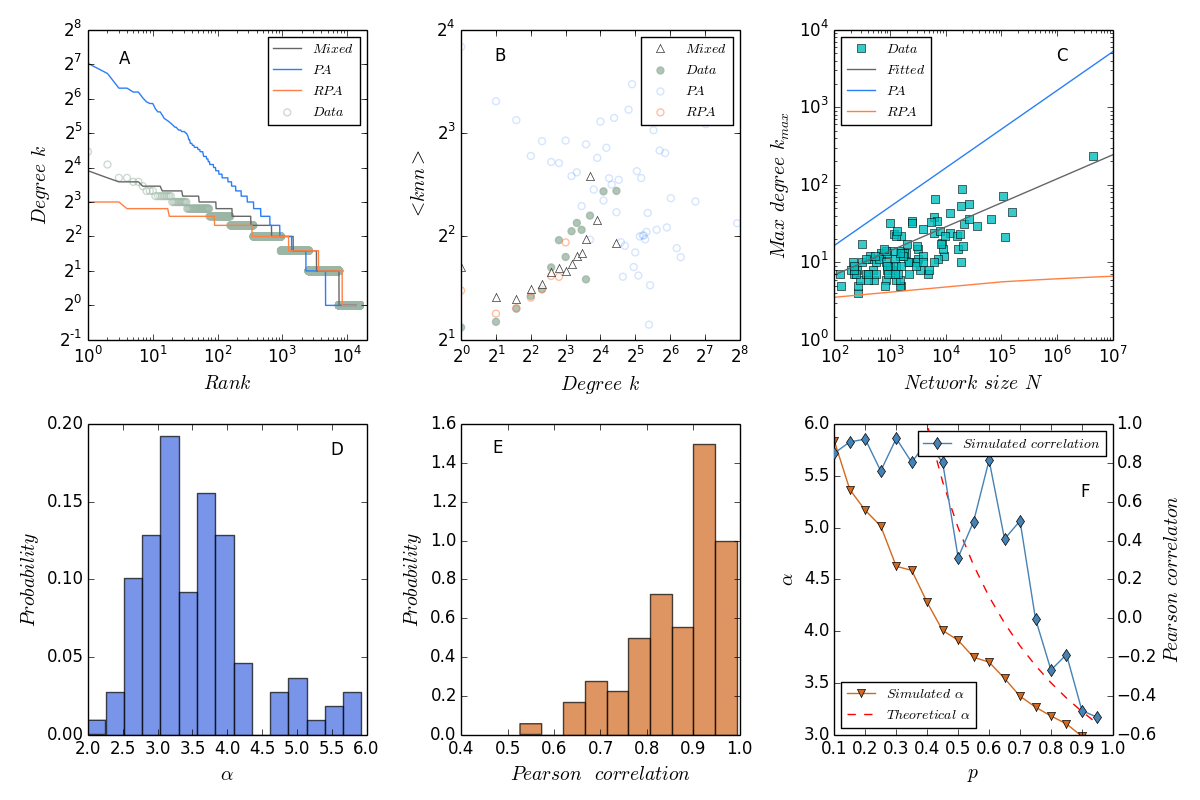}}
\caption{Comparing different models against the empirical observations. In Panel A we show the the rank-order curve of degree (log degree vs. log rank) of the Physics network after the evolution over 5.5 years (gray circles), which is compared against the corresponding curves generated by three models, including the PA model (blue line), the RPA model (orange line) and the mixing model with a $p$ equals 0.3 (black line). It is observed that the mixing model fits the data better than the other two models. Panel B shows the change of neighbor connectivity $<k_{nn}>$ with the increase of degree $k$ in the Physics network (gray dots) and in three models. The PA model (blue circles) does not give the positive correlation between these two variables. The RPA model (orange circles) can not reproduce the behavior of high-degree nodes. Again, the mixing model (black triangles) does better than the other two models. In Panel C we show the relationship between highest degree $k_{max}$ and network size $N$ across 110 networks. Each data point represents a network. The empirical data points lay between the upper bound given by the PA model, which is $k_{max}\sim N^{0.5}$ (blue line), and the lower bound given by the RPA model, which is approximately $k_{max}\sim ln(N)^{0.5}$ (orange line). The OLS regression fitting in the log-log axes gives  $k_{max}\sim N^{0.3}$ (black line). Panel D and E show the distribution of power-law exponent $\alpha$ of degree distribution fitted using the maximum likelihood method [23] and the Pearson correlation $r$ between $k_{nn}$ and $k$.  Panel F shows that the mixing model reproduces different values of $\alpha$ (brown triangels) and Pearson correlation $r$ (blue diamonds) by changing the value of $p$ in simulation. We also plot the theoretical relationship $\alpha~1+2/p$ (red line) to guide the eye. The deviation of the simulated data (brown triangles) from the theoretical line (red line) is due to the limited system size effect in simulation.
}
\label{figure3}
\end{center}
\end{figure*}

In Panel A and B in Figure \ref{figure3}, we analyze the Physics network after the evolution of 5.5 years. It is observed that the rank-order curve of degree (log degree vs. log rank) lies between Zipf's law of exponent $0.5$ (which equals to a power-law distribution of an exponent equals $3$ according to \cite{adamic2000zipf,}) predicted by the PA model and the curve predicted by the RPA model. The highest degree, i.e., the maximum number of answers collected by a single question, is only 22 among all the 15,717 nodes. By fitting the simulated mixing networks to data, we estimate that $p=0.3$. We find that the simulated network with parameter $p=0.3$ also successfully reproduces the assortative mixing pattern of the Physics network, whereas the PA model does not show assortativity and the RPA model can not replicate the behavior of high-degree nodes. 

A systematic investigation shows that a majority of the studied 110 networks exhibit two properties: 1) the mitigation of degree inequality among nodes, which is characterized by power-law distributions with an exponent $\alpha \leq 3$ (Figure \ref{figure3}A), and, as a result, the slow increase of highest degree with network size (Figure \ref{figure3}C); 2) the assortative mixing of nodes (Figure \ref{figure3}E). In Figure \ref{figure3}F we show than both of these two properties can be reproduced by the mixing model. In particular, by tuning the value of $p$ we obtain networks of different levels of degree inequality and assortativity. Generally, when we increase the parameter $p$ in the mixing model, the degree distribution becomes more uneven and the assortative mixing property of the system disappears.

\section*{Conclusion and Discussion}

To obtain understanding into the mechanisms underlying the creation of knowledge, we analyze the data of a very large online Q\&A system that includes the question asking and answering activities of millions of users over five years. We find a separation between old, popular knowledge and new, less popular knowledge, which can be explained by two different attention allocation mechanisms in the system. We combine these mechanisms in a mixing model and successfully reproduce the observed properties of knowledge networks.

The assumption on the existence of two attention allocation mechanisms is not only interesting by itself, but also provide insights for community management. The current reputation point system of StackExchange has been criticized a lot \cite{tausczik2012participation}, especially that it emphasizes more on the quantity of contribution than the quality of contribution. We argue that the current policy that emphasizes the number of contributions actually motivate users to answer more easy, new questions to earn reputation points, thus migrates the concentration of attention on popular questions and makes sure new questions are addressed promptly. However, the existence of two attention allocation mechanisms, which may due to the existence and division of two types users who favor and dislike competition in answering questions respectively, has a negative impact on the integration of knowledge. Therefore, if StackExchange can modify the reputation system to encourage the mixing behaviors, namely, answering both old and new questions, that may bridge the separation between knowledge created at different time points and accelerate the adaption and diffusion of novel ideas.

\section*{Author contributions}

L. W. and M.A.J. designed research; L.W. analyzed data and did the analytical work; L.W., and M.A.J. wrote the paper.

\section*{Acknowledgments}

We acknowledge financial support for this work from the National Science Foundation, grant number 1210856. L. W. thanks Yazhu Song for her inspiring discussions. 

\bibliography{attensionDynamics}

\providecommand{\noopsort}[1]{}\providecommand{\singleletter}[1]{#1}%
\begin{thebibliography}{10}
\providecommand{\url}[1]{\texttt{#1}}
\providecommand{\urlprefix}{URL }
\expandafter\ifx\csname urlstyle\endcsname\relax
  \providecommand{\doi}[1]{doi:\discretionary{}{}{}#1}\else
  \providecommand{\doi}{doi:\discretionary{}{}{}\begingroup
  \urlstyle{rm}\Url}\fi
\providecommand{\bibAnnoteFile}[1]{%
  \IfFileExists{#1}{\begin{quotation}\noindent\textsc{Key:} #1\\
  \textsc{Annotation:}\ \input{#1}\end{quotation}}{}}
\providecommand{\bibAnnote}[2]{%
  \begin{quotation}\noindent\textsc{Key:} #1\\
  \textsc{Annotation:}\ #2\end{quotation}}
\providecommand{\eprint}[2][]{\url{#2}}

\bibitem{simon1973applying}
Simon HA (1973) Applying information technology to organization design.
\newblock Public Administration Review 33: 268--278.
\bibAnnoteFile{simon1973applying}

\bibitem{wu2007novelty}
Wu F, Huberman BA (2007) Novelty and collective attention.
\newblock Proceedings of the National Academy of Sciences 104: 17599--17601.
\bibAnnoteFile{wu2007novelty}

\bibitem{weng2012competition}
Weng L, Flammini A, Vespignani A, Menczer F (2012) Competition among memes in a
  world with limited attention.
\newblock Scientific Reports 2.
\bibAnnoteFile{weng2012competition}

\bibitem{lehmann2012dynamical}
Lehmann J, Gon{\c{c}}alves B, Ramasco JJ, Cattuto C (2012) Dynamical classes of
  collective attention in twitter.
\newblock In: Proceedings of the 21st international conference on World Wide
  Web. ACM, pp. 251--260.
\bibAnnoteFile{lehmann2012dynamical}

\bibitem{wu2013metabolism}
Wu L, Zhang J, Min Z (2014) The metabolism and growth of web forums.
\newblock PloS one 8: e102646.
\bibAnnoteFile{wu2013metabolism}

\bibitem{wu2013decentralized}
Wu L, Zhang J (2013) The decentralized flow structure of clickstreams on the
  web.
\newblock The European Physical Journal B 86: 1--6.
\bibAnnoteFile{wu2013decentralized}

\bibitem{barabasi1999emergence}
Barab{\'a}si AL, Albert R (1999) Emergence of scaling in random networks.
\newblock science 286: 509--512.
\bibAnnoteFile{barabasi1999emergence}

\bibitem{pastor2001dynamical}
Pastor-Satorras R, V{\'a}zquez A, Vespignani A (2001) Dynamical and correlation
  properties of the internet.
\newblock Physical review letters 87: 258701.
\bibAnnoteFile{pastor2001dynamical}

\bibitem{newman2002assortative}
Newman ME (2002) Assortative mixing in networks.
\newblock Physical review letters 89: 208701.
\bibAnnoteFile{newman2002assortative}

\bibitem{wu2015hidden}
Wu L, Wang CJ, Janssen M, Zhang J, Zhao M (2015) The hidden geometry of
  attention diffusion.
\newblock arXiv preprint arXiv:150106552 .
\bibAnnoteFile{wu2015hidden}

\bibitem{kobourov2012spring}
Kobourov SG (2012) Spring embedders and force directed graph drawing
  algorithms.
\newblock arXiv preprint arXiv:12013011 .
\bibAnnoteFile{kobourov2012spring}

\bibitem{guan2014fine}
Guan Z, Yang S, Sun H, Srivatsa M, Yan X (2014) Fine-grained knowledge sharing
  in collaborative environments.
\newblock IEEE Transactions on Knowledge and Data Engineering 27: 2163.
\bibAnnoteFile{guan2014fine}

\bibitem{dunne2002food}
Dunne JA, Williams RJ, Martinez ND (2002) Food-web structure and network
  theory: the role of connectance and size.
\newblock Proceedings of the National Academy of Sciences 99: 12917--12922.
\bibAnnoteFile{dunne2002food}

\bibitem{sevim2006effects}
Sevim V, Rikvold PA (2006) Effects of preference for attachment to low-degree
  nodes on the degree distributions of a growing directed network and a simple
  food-web model.
\newblock Physical Review E 73: 056115.
\bibAnnoteFile{sevim2006effects}

\bibitem{mislove2013empirical}
Mislove A, Koppula HS, Gummadi KP, Druschel P, Bhattacharjee B (2013) An
  empirical validation of growth models for complex networks.
\newblock In: Dynamics On and Of Complex Networks, Volume 2, Springer. pp.
  19--40.
\bibAnnoteFile{mislove2013empirical}

\bibitem{eisenberg2003preferential}
Eisenberg E, Levanon EY (2003) Preferential attachment in the protein network
  evolution.
\newblock Physical review letters 91: 138701.
\bibAnnoteFile{eisenberg2003preferential}

\bibitem{capocci2006preferential}
Capocci A, Servedio VD, Colaiori F, Buriol LS, Donato D, et~al. (2006)
  Preferential attachment in the growth of social networks: The internet
  encyclopedia wikipedia.
\newblock Physical Review E 74: 036116.
\bibAnnoteFile{capocci2006preferential}

\bibitem{bianconi2001competition}
Bianconi G, Barab{\'a}si AL (2001) Competition and multiscaling in evolving
  networks.
\newblock EPL (Europhysics Letters) 54: 436.
\bibAnnoteFile{bianconi2001competition}

\bibitem{wang2013quantifying}
Wang D, Song C, Barab{\'a}si AL (2013) Quantifying long-term scientific impact.
\newblock Science 342: 127--132.
\bibAnnoteFile{wang2013quantifying}

\bibitem{barrat2005rate}
Barrat A, Pastor-Satorras R (2005) Rate equation approach for correlations in
  growing network models.
\newblock Physical Review E 71: 036127.
\bibAnnoteFile{barrat2005rate}

\bibitem{adamic2000zipf}
Adamic LA (2000) Zipf, power-laws, and pareto-a ranking tutorial.
\newblock Xerox Palo Alto Research Center, Palo Alto, CA, http://ginger hpl hp
  com/shl/papers/ranking/ranking html .
\bibAnnoteFile{adamic2000zipf}

\bibitem{tausczik2012participation}
Tausczik YR, Pennebaker JW (2012) Participation in an online mathematics
  community: differentiating motivations to add.
\newblock In: Proceedings of the ACM 2012 conference on Computer Supported
  Cooperative Work. ACM, pp. 207--216.
\bibAnnoteFile{tausczik2012participation}

\end{thebibliography}

\end{document}